\documentclass[useAMS,usenatbib,usegraphicx]{mn2e}

\title[Gamma-rays from Cluster-scale AGN Outbursts]{Gamma-ray emission associated with Cluster-scale AGN Outbursts}
\author[Hinton, Domainko \& Pope]{J. A. Hinton$^{1}$\thanks{E-mail: j.a.hinton@leeds.ac.uk}, W. Domainko$^{2}$ \& E. C. D. Pope$^{1}$\\
$^{1}$School of Physics \& Astronomy, University of Leeds, Leeds LS2 9JT, UK\\
$^{2}$Max-Planck-Institut f\"ur Kernphysik, Heidelberg, D 69029, Germany}

\begin{document}


\pagerange{\pageref{firstpage}--\pageref{lastpage}} \pubyear{2007}

\maketitle

\label{firstpage}

\begin{abstract}
Recent observations have revealed the existence of enormously
energetic $\sim 10^{61}$ erg AGN outbursts in three relatively distant
galaxy clusters. These outbursts have produced bubbles in the
intra-cluster medium, apparently supported by pressure from
relativistic particles and/or magnetic fields.  Here we argue that if
$\geq$GeV particles are responsible then these particles are very
likely protons and nuclei, rather than electrons, and that the
$\gamma$-ray emission from these objects, arising from the
interactions of these hadrons in the intra-cluster medium, may be
marginally detectable with instruments such as GLAST and HESS.
\end{abstract}


\section{Introduction}

Galaxy clusters are the largest gravitationally bound systems in the
universe. In addition, radio \citep{giovannini00, feretti04}
and hard X-ray \citep{rephaeli02, fusco04}) observations have revealed that a significant
component of non-thermal particles can be found in such systems. The
remaining tracer of non-thermal particles is high energy $\gamma$-ray
emission, but no such signal has been firmly detected from galaxy
clusters so far \citep{reimer03}.

Despite this non-detection a number of arguments suggest that galaxy
clusters are potentially powerful emitters of high energy radiation.
\citet{voelk96} and \citet{berezinsky97} recognised that hadronic cosmic rays (CRs)
with energies of less than 10$^{15}$ eV accumulate within the cluster
volume for the entire Hubble time. This CR component, together with
the presence of target material in the form of the hot intra-cluster
medium (ICM), will lead to very high energy (VHE) $\gamma$-ray
production via inelastic proton-proton collisions and subsequent
$\pi^0$ decay \citep{dennison80,voelk96}. Furthermore, leptonic CRs are also capable of
generating high energy electromagnetic radiation: TeV electrons may
up-scatter cosmic microwave background (CMB) photons to $\gamma$-ray
energies in the inverse Compton processes \citep{atoyan00, gabici03,gabici04}. 
The lifetime of VHE electrons, however, is limited
by IC and synchrotron losses to $\sim 10^{6}$ years (for typical
cluster magnetic field strengths).  Therefore, only recently injected
electrons will contribute to the production of VHE
$\gamma$-rays. Finally, if populations of ultra high energy
($>10^{18}$ eV) CR protons exist in galaxy clusters, they will
interact with CMB photons and produce electron - positron pairs which
will in turn radiate TeV photons (with a characteristically hard
energy spectrum) via the IC mechanism \citep{inoue05}.

Several sources of CRs are plausible in galaxy clusters. Large scale
shock waves caused by hierarchical structure formation may accelerate
particles to sufficiently high energies 
\citep{colafrancesco00,loeb00,ryu03}. Additionally supernovae and galactic winds from
cluster galaxies can populate galaxy clusters with non-thermal
particles \citep{voelk96}. Furthermore, powerful AGNs
are believed to be prominent injectors of CRs into the ICM 
\citep{ensslin97,aharonian02,pfrommer04}.

In this paper we focus on the scenario where a powerful AGN injects
high energy particles into a galaxy cluster. Prominent AGNs are often
found in galaxy clusters with short central cooling times of the
thermal ICM \citep[for a review of these so called \emph{cooling flow
clusters} see][]{fabian94} and their effect on the ICM can
be seen in several systems \citep[see][for a
sample of such galaxy clusters]{birzan04}.  High resolution X-ray observations
have revealed bubbles, cavities and weak shocks in the ICM driven by
activity of the central galaxy in several systems (e.g.
\citealt{boehringer93,blanton01,schindler01,mcnamara01,fabian03,choi04}). 
Bubbles in the
X-ray gas are often associated with radio lobes, indicating the
presence of relativistic electrons \citep{owen00,fabian02,gitti06}.  Very recently,
so-called \emph{cluster-scale} AGN outbursts have been found in three
clusters, all with estimated mechanical energy of at least 10$^{61}$
erg. These systems are: MS\,0735.6+7421 
\citep{mcnamara05}, Hercules A \citep{nulsen05a}
and Hydra A \citep{nulsen05b,wise07}.  Large scale radio emission was found in Hydra A
with deep VLA observations \citep{lane04}. Due to the
large amount of energy input into these systems by the central AGN,
plausibly in the form of relativistic particles, these systems may be
promising targets for high (and very high) energy $\gamma$-ray
observations.

Experimentally, $\gamma$-ray astronomy is in a phase of rapid
development. Several Imaging Atmospheric Cherenkov Telescopes (IACTs)
have recently been completed: HESS \citep{hinton04},
MAGIC \citep{lorenz04} and VERITAS \citep{krennrich04}). The combination of these VHE instruments with the
high energy (HE) detector GLAST \citep{thompson04}, due for
launch early next year, will provide sensitive coverage of the 100~MeV
to 10~TeV energy regime for the first time. Here we will argue that
these instruments may be close to the sensitivity threshold required
to detect the high energy electromagnetic signatures of large scale
AGN outbursts in galaxy clusters.

In this paper we investigate the high energy luminosity of the three
known galaxy clusters which host cluster-scale AGN outbursts. The
relevant properties of these systems are given in Table~\ref{table1}.

\begin{table*}
 \centering
 \begin{minipage}{140mm}
  \caption{Characteristics of powerful AGN outbursts. The flux $F^{\prime}_{\nu}$ is calculated under the assumption of \emph{scenario A} (discussed later).}
  \begin{tabular}{lcccccc}
  \hline
Object   &       $z$   &    $PV$     &     Age    &  mean Density $n_{e}$  &  Bubble Diam. &  $\nu F^{\prime}_{\nu}$ (100 GeV) \\
         &             &  ($10^{61}$ erg) &   ($10^{8}$  years) &  ($10^{-3}$ cm$^{-3}$) &    (kpc/$''$) & ($10^{-13}$ erg cm$^{-2}$ s$^{-1}$)\\ \hline
MS\,0735.6+7421  & 0.22   &     6    &      1.0    &          3    &        240/70      &     0.7 \\
Hercules A     & 0.154  &     3      &      0.6      &          5    &        160/55      &     1.2 \\
Hydra A        & 0.0538 &    0.9/0.41 &      1.4      &          5    &        210/200     &     3.0 \\
\hline
\end{tabular}
\end{minipage}
\label{table1}
\end{table*}

\section{Cosmic rays in Cluster-scale Outbursts}

The existence of radio synchrotron emission in coincidence with X-ray
cavities seen in AGN outbursts indicates the existence of relativistic
electrons within the bubbles and leads naturally to the suggestion
that relativistic particles may support these cavities.  This
hypothesis must be confronted with three key issues: 1. can
relativistic particles be confined for the required timescales?
2. are the energy loss timescales of these particles sufficiently
long?  3. is there observational evidence for other contributions to
the bubble pressure?

\begin{figure}
\centering
\vspace{2mm}
\includegraphics[width=9.2cm]{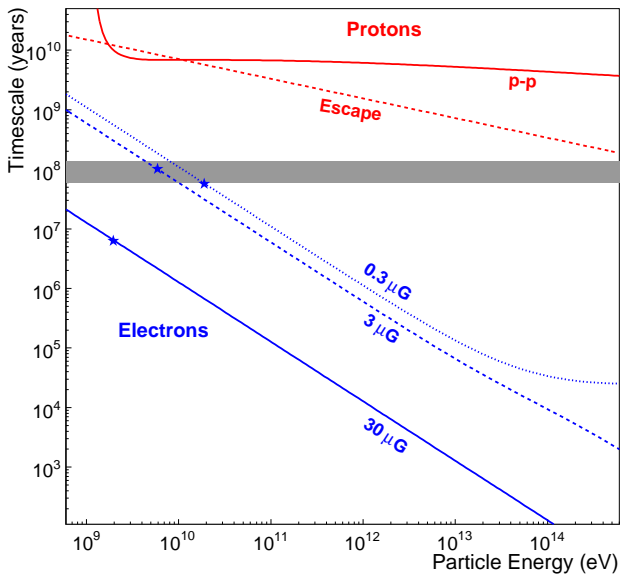}
\vspace{-3mm}
\caption{ Timescales of relevance to AGN outbursts. For protons the
  mean time between inelastic collisions (solid line) and the escape
  time from the system (dashed line, assuming diffusion as given by
  eq. 9 of V\"olk et al~(1996) with $B$ = 3 $\mu$G), are shown. Note
  that this is the escape time from a 100~kpc radius bubble, the time
  required to escape the cluster (under the same assumptions) is $\sim
  10^{10}$ years for a $10^{15}$ eV proton. For electrons the energy
  loss timescale ($E/\frac{dE}{dt}$) due to synchrotron and inverse
  Compton radiation is shown for three different magnetic field
  strengths. The stars show the mean electron energy contributing to
  1~GHz synchrotron emission for the three different field strengths.
  An ambient density of $5\times10^{-3}$ protons cm$^{-3}$ is
  assumed. The target photon field for inverse Compton scattering is
  taken solely as the CMB. This is then an upper limit on the cooling
  timescale, since infrared photons from the central galaxy may also
  act as targets for the inverse Compton process. The shaded band
  indicates the apparent ages of the outbursts considered here.
}
\label{f1}
\end{figure}

For cluster-scale outbursts the timescales involved have been
estimated at $\sim 10^{8}$ years, considerably longer than for
previously identified systems \citep{birzan04}. Fig~\ref{f1} shows the relevant energy loss
timescales for ultrarelativistic electrons in the central cluster
environment.  We find that for the typical $B$-fields of a few $\mu$G
found in the central regions of clusters, TeV electrons lose their
energy on timescales of $10^{6}$ years via inverse Compton and
synchrotron cooling (see Fig~\ref{f1}). Only $<$ GeV electrons can
exist at the outer edge of a bubble of age $\sim 10^{8}$ years. The
observed radio synchrotron emission should therefore provide evidence
for spectral cooling away from the central AGN. In the case of Hydra A
this spectral steepening is clearly seen \citep{lane04}. The measurements are broadly consistent with the
injection of a power-law of electrons ($dN/dE \propto E^{-\alpha}$)
with $\alpha \sim 2$
which is subsequently cooled above a critical energy
$E_{\mathrm{crit}} \sim 100$ MeV to $\alpha \sim 3$ (resulting in
radio synchrotron spectral index of $\sim 1$).  By considering
continuous injection and synchrotron cooling of the electrons over the
lifetime of the outburst, we estimate that the currently radio
emitting electrons represent roughly 1/3 of the total energy injected
in electrons over this period. To estimate the energy carried by these
electrons an estimate of the magnetic field strength inside the
bubbles is required. Rotation measure estimates of the magnetic field
\emph{outside} the radio lobes of Hydra A suggest $B \sim$ 30 $\mu$G
\citep{taylor93}.  If the B-field within the lobes is
similar then the total energy in electrons, and in the magnetic field
itself, are both close to $10^{59}$ erg. As the total energy
associated with the Hydra A outburst is $\sim 10^{61}$ erg it seems
that at least in this system there is missing pressure within the
bubbles unless the magnetic field strength there is $\gg 30 \mu$G. To
support the Hydra A bubbles solely with magnetic pressure requires $B
\sim 300 \mu$G. Such high fields seem unlikely due to the observation
of smooth features in the 1.4~GHz emission on $\sim 100$ kpc
scales. The synchrotron cooling time of the 1.4 GHz emitting (500 MeV)
electrons in a 300 $\mu$G field is such that rectilinear propagation
at speeds very close to $c$ would be required to reach the edge of these
features.  The missing pressure may be ascribed to thermal gas \citep{gitti07}
or hadronic cosmic rays \citep{dunn04}. We note that in the past proton dominated jets have
been suggested \citep{celotti93, sikora00} injecting hadronic cosmic rays into the ICM during
AGN outbursts. Cosmic ray protons and nuclei have much less severe
energy losses than electrons.  For a proton to electron ratio of only
1/30 these particles could provide the energy required to support the
bubbles.  Furthermore, the escape time of $<10^{15}$ eV protons from
the central 200~kpc of these clusters is likely to be longer than the
lifetime of the bubbles \citep{voelk96}. A large
fraction of the injected CRs may therefore be confined within the
observed bubbles for the required $10^{8}$ years (see Fig.~\ref{f1})
but the particles with the highest energies are likely able to
penetrate the thermal gas which surrounds the bubbles (see
Fig.~\ref{f2}).  We further note that in cluster scale AGN outbursts
the central black hole must accrete a significant fraction of its own
mass in a single activity period and convert it very efficiently into
mechanical power \citep{mcnamara05, nulsen05a, nulsen05b}. Therefore there
is not much room for additional radiative energy losses of the
relativistic particles driving the observed shock fronts.

For Hydra-A it therefore seems likely that thermal gas or hadronic CRs
dominate the bubble pressure\footnote{We note that this conclusion
rests on the extrapolation of the observed radio spectrum in Hydra-A
down to $<$ 400 MeV electron energies (i.e. radio frequencies below 74
MHz).  Electrons of a few hundred MeV have long enough lifetimes to
support the bubbles, and could have sufficient energy if there is a
second spectral component at these energies}. For the other objects
considered here such detailed radio observations do not exist. In
general magnetic fields and/or low energy electrons can be considered
viable alternatives as the dominant energy content of AGN jets (see
for example \citealt{deyoung06} and \citealt{dunn06}).

Assuming that hadronic cosmic rays are primarily responsible for the
bubble expansion, two additional inputs are needed to calculate the
$\gamma$-ray flux from these objects. Firstly, an estimate of the
internal energy in CRs at the present day and secondly an estimate of
the target density in and around the bubbles. Both these numbers
require consideration of the hydrodynamics of the outbursts, which are
discussed in the next section.

\section{Bubble properties}

From the X-ray data we can determine, or constrain, many properties of
the AGN-inflated bubbles. Observations of the X-ray surface brightness
profile yield the pressure of the gas outside the bubble, the bubble
volume, and limits to the emissivity of the material inside the
bubble. This evidence indicates that the density of bubble material is
low compared to its surroundings.

It is probably reasonable to assume an approximate pressure balance
between the material inside the bubble and the ambient gas. Therefore,
the external pressure is balanced by the total pressure exerted by the
sum of the partial pressures of each component in the bubble: thermal
pressure ($P_{\rm th}$), cosmic ray pressure ($P_{\rm CR}$), magnetic
pressure ($P_{\rm mag}$) and the relativistic electron pressure
($P_{\rm elec}$). However, the radio data indicate that, in the case
of Hydra-A, the magnetic and electron partial pressures are likely
less significant than the thermal and cosmic ray
components. Henceforth, we will consider the bubble to be a mixture of
only thermal material and cosmic rays and define $X_{\rm CR} = P_{\rm CR}/P_{\rm
ext}$, where $P_{\rm ext}$ is the external pressure on the bubble.   
In this case, the total internal energy content within the bubble is,
\begin{equation}
  E_{\rm int} = V\bigg(\frac{P_{\rm th}}{\gamma_{\rm th} -1} + \frac{P_{\rm CR}}{\gamma_{\rm CR} -1}\bigg),
\end{equation}
where $\gamma_{\rm th} = 5/3$ is the adiabatic index of the thermal
material, and $\gamma_{\rm CR}= 4/3$ is the adiabatic index of the
relativistic cosmic rays. In this description, the density inside the
bubble is governed by $P_{\rm CR}$, and the temperature of the
material. We note that the case of over-pressurised
bubbles $X_{\rm CR}$ may be greater than 1.

Another useful parameter when discussing the properties of bubbles, is
the density contrast ($\eta$) between the material within the bubble
$\rho_{\rm bub}$ and the ambient material $\rho_{\rm ext}$, $\eta
=\rho_{\rm ext}/\rho_{\rm bub} $. It should be noted that the cosmic
rays make a negligible contribution to $\rho_{\rm bub}$. If we
consider a jet that injects CRs and hot gas at a constant rate, then
if this inflates a bubble, the average density within the bubble is,
\begin{equation} \label{eq:rb}
  \rho_{\rm bub} = \frac{\dot{m}t}{V(t)}
\end{equation}
where $\dot{m}$ is the mass injection rate of the jet, and $V(t)$ is
the time-dependent volume of the cavity. The enthalpy of a slowly
inflated bubble is,
\begin{equation}\label{eq:enth}
  E = L_{\rm jet}t = \frac{\gamma}{\gamma -1}P_{\rm ext}V(t)
\end{equation}
where $L_{\rm jet}$ is the constant jet power and $\gamma$ is the
effective adiabatic index, such that
\begin{equation}\label{eq:gamma}
\frac{\gamma}{\gamma -1} = \frac{\gamma_{\rm th}}{\gamma_{\rm th} -1}(1-X_{\rm CR}) + \frac{\gamma_{\rm CR}}{\gamma_{\rm CR} -1} X_{\rm CR} 
\end{equation}
Therefore, we see that the volume of the bubble is proportional to the
duration of the AGN outburst.  Consequently, since the mass injection
rate is also constant, so must be $\rho_{\rm bub}$. Combining
equations (\ref{eq:rb}) and (\ref{eq:enth}) and substituting $P_{\rm
ext} = k_{\rm b}T_{\rm ext}\rho_{\rm ext}/\mu$ gives the density
contrast,
\begin{equation}
  \eta = \frac{\gamma-1}{\gamma}\frac{\mu L_{\rm jet}} {\dot{m}k_{\rm b}T_{\rm ext}}
\end{equation}
which depends on the jet parameters, and the temperature of the
ambient gas.  Assuming a constant mass injection rate, $\eta$
increases for greater jet power because the density of material within
the bubble is lower.  $\eta$ is roughly constant (and $\gg 1$) whilst
the jet is active.  After the jet switches off and the bubble rises
buoyantly, it will expand to maintain pressure equilibrium with its
surroundings, and the density contrast will fall. We can obtain a
simple description of this by assuming that the bubble behaves
adiabatically.  Under adiabatic conditions, the specific entropy of
the bubble will remain constant, thus,
\begin{equation}
  \frac{P_{\rm bub}}{\rho_{\rm bub}^{\gamma}} = \rm constant.
\end{equation}
Since the temperature of the ICM is roughly constant with radius,
$P_{\rm ext} \propto \rho_{\rm ext}$, and the bubble is in pressure
equilibrium with its surroundings, we have: $\rho_{\rm bub} \propto
\rho_{\rm ext}^{1/\gamma}$, and $\eta \propto \rho_{\rm
ext}^{1-1/\gamma}$. Thus, the density of the bubble falls less steeply
than that of the ICM.

\begin{figure*}[ht]
\centering
\vspace{-1mm}
\includegraphics[width=18cm]{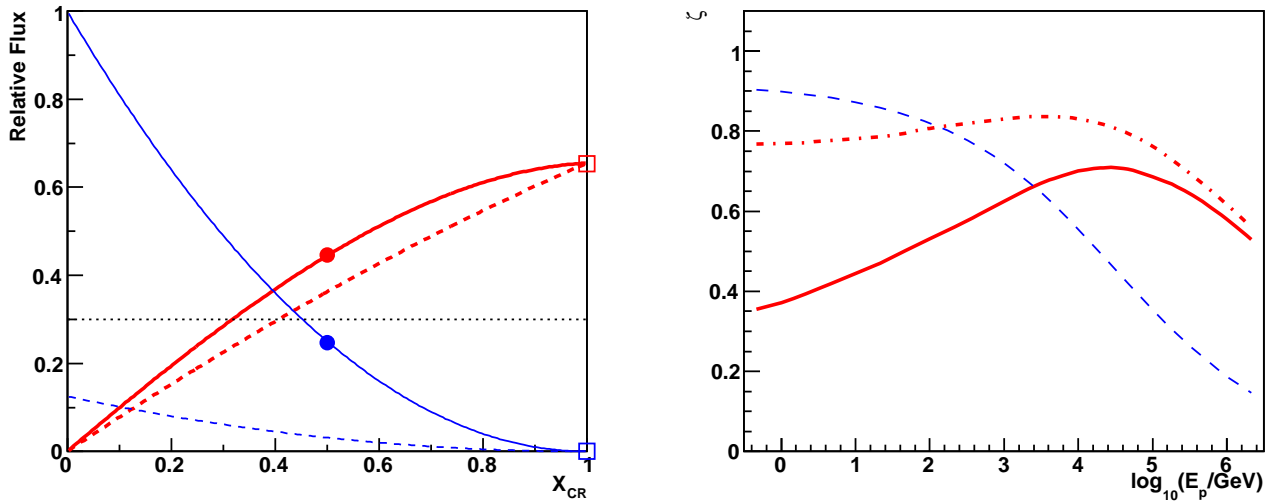}
\vspace{-4mm}
\caption{ Left: Relative flux of X-rays and $\gamma$-rays as a
function of the relative CR pressure $X_{CR}$. The X-ray flux from
inside the bubbles (thin falling lines) is expressed relative to that
of an equal volume of gas at the temperature and pressure of the
external medium. The $\gamma$-ray flux from the system is shown
relative to that expected for a density contrast $\eta=1$. Two cases
are shown: $T_{\mathrm{int}} = T_{\mathrm{ext}}$ (solid line) and
$T_{\mathrm{int}} = 4 T_{\mathrm{ext}}$ (dashed line). The dotted horizontal line
indicates the approximate range of X-ray contrast excluded by
observations.  Right: The energy dependence of CR confinement within
the bubbles (dashed line) and the mixing parameter $\zeta$ for two
assumptions for $X_{CR}$ and $T_{\mathrm{int}}$, $X_{CR}=0.5$,
$T_{\mathrm{int}}=T_{\mathrm{ext}}$ (dash-dotted curve, indicated by
solid circles in the left panel) and $X_{CR}=1.0$ (solid curve,
indicated by open squares in the left panel. In both panels the system
geometry, external density profile and diffusion coefficient are
specific to the Hydra A system. Low energy particles do not penetrate
far into the external medium and hence encounter mostly low density
gas inside the bubble.  At the highest energies CRs begin to escape
from the high density parts of the cluster into the more tenuous
ICM. Note that the CRs with log$_{10}(E/\mathrm{GeV}) = 4$ are those
primarily responsible for the 1 TeV $\gamma$-ray emission.  }
\label{f2}
\end{figure*}

The argument above describes the slowest possible rate at which $\eta$
can approach unity. In practise, bubbles probably behave far from
adiabatically.  Analysis of recent hydrodynamic simulations using the
FLASH code, suggests that the Kutta-Zhukovsky force plays an important
part in mixing the bubble with the ICM and dissipating the bubble
enthalpy over a relatively short timescale
\citep{pavlovski07}
\footnote{
This force acts to expand the bubble radially, but perpendicularly to the 
direction of ascent. As the bubble expands an additional component of the 
Kutta-Zhukovsky force pushes the bubble down, reducing its ascent 
velocity \citep{landau95}.
}. These simulations show that $\eta$ approaches a
value of 1, from an initial value of 1000, after rising only a few
bubble radii. The observed X-ray contrast of the bubbles can be used
to place limits on the density inside the bubbles for a given 
temperature. The X-ray contrast is approximately
$(n_{\mathrm{int}}/n_{\mathrm{ext}})^{2}
(T_{\mathrm{int}}/T_{\mathrm{ext}})^{1/2}$.  The uncertainties on the
X-ray measurements and the geometry of the bubbles are such that this
quantity must be less than about $\sim 1/3$ for the Hydra A system
\citep{wise07}. Fig.~\ref{f2} shows this X-ray contrast
as a function of the relative CR pressure $X_{CR}$ for two different
assumed temperatures $T_{\mathrm{int}}$.

In the context of this work, it is also important to know the density
of material in the bubble rim. In the early stages, the bubble
inflation is likely to be highly supersonic, meaning that the density
of the rim material should be roughly four times denser than the
ambient material. Since the emissivity is proportional to the square
of the gas density, this will lead to the appearance of bright rims
surrounding bubbles. However, because the cooling time of these rims
is short compared to that of the surrounding gas, the rims are
expected (and observed, see \citealt{blanton04} and \citealt{gitti07}) to be cold ($kT \ll 1$ keV). Using the cooling
function given by \citet{sutherland93} we find
that the time taken to cool from 4 keV (the approximate temperature of
the gas 100 kpc from Hydra-A) to $\ll 1$ keV is: $t_{\mathrm{cool}}
\approx 1.4 \times 10^{8} (n_{0}/10^{-2} \mathrm{cm}^{-3})^{-1}
(kT_{0}/\mathrm{4 keV})^{1.5} \mathrm{years}$.  Assuming a compression
ratio of four, the rims in these systems would have a density $2
\times 10^{-2} \mathrm{cm}^{-3}$, and hence cool in about half the age
of the observed bubbles.

\section{Model}
\label{sec_model}

As discussed in the previous sections we assume here that hadronic
cosmic rays are responsible for expanding the observed bubbles.  To
calculate the rate of proton-proton interactions, and the secondary
particle production in these interactions, we apply the
parameterisations derived by \citet{kelner06} based on the
SIBYLL hadronic interaction model \citep{fletcher}.
Adapting \citet{kelner06} (eq. 69), the flux of
$\gamma$-rays from proton-proton (and nucleus-nucleus) interactions
can be expressed as:

\begin{eqnarray}\label{eq:gammaflux}
\frac{dN_{\gamma}}{dE_{\gamma}} &=& \kappa n_{e}cV\epsilon(E_{\gamma})\nonumber\\
&\times&\int_{E_{\gamma}}^{\infty}\zeta(E_{p})\sigma_{pp}(E_{p})\frac{dn_{p}}{dE_{p}} 
f_{\gamma}(E_{\gamma}/E_{p},E_{p})\frac{dE_{p}}{E_{p}}
\end{eqnarray}

$\kappa n_{e}$ is the effective cross-section weighted number density
of target (thermal) protons and nuclei (a thin target approximation is
appropriate as $t_{pp} >> t_{bubble}$).  Assuming a primordial
abundance of both CRs and target nuclei we find $\kappa=1.15$. $c$ is
the speed of light, $V$ is the bubble volume, $\epsilon(E_{\gamma})$
represents the $\gamma$-$\gamma$ absorption on the (IR) extragalactic
background light (EBL). We note that following recent constraints from
the HESS \citep{aharonian06} and Spitzer \citet{dole06} instruments, uncertainties on the EBL in the relevant
wavelength range are now greatly reduced. Here we calculated EBL
absorption $\epsilon(E_{\gamma})$ using the wavelength dependant
($z=0$) EBL density given in Fig. 13 of \citet{dole06},
ignoring evolutionary effects. The function $\zeta(E_{p})$ reflects
the degree of mixing between the target nuclei and the CRs. $\zeta=1$
corresponds to the case of CRs confined in a uniform medium with
electron density $n_{e}$.  $f_{\gamma}(E_{\gamma}/E_{p},E_{p})$ is the
energy distribution function for $\gamma$-rays produced in an average
interaction of a proton of energy $E_{p}$ \citep[see equation 56 of][]{kelner06}.
The spectral energy density of protons is assumed to be of the form:

\begin{equation}\label{eq:crspec}
\frac{dn_{p}}{dE_{p}} = k E_{p}^{-\alpha}\exp{\frac{-E_{p}}{E_{\max}}}
\end{equation}

With $k$ chosen such that $V \int_{\mathrm{1 GeV}}^{\infty}
dn_{p}/dE_{p} = E_{\mathrm{CR}} = 3P_{\mathrm{ext}}V
X_{\mathrm{CR}}$. Diffusive shock acceleration, either relativistic
\citep[see for example][]{kirk00} or non-relativistic,
predicts a spectral index of injected particles close to 2.  We
therefore set $\alpha=2$ for the purpose of our calculations.  The
predicted peak flux (in the GeV range) varies only weakly with the
assumed energy range.  In contrast, the predicted TeV emission depends
strongly on the maximum energy of the injected particles. The value of
at least $E_{\mathrm{max}} = 10^{14}$ eV (as assumed here) is required
to produce up to 10 TeV $\gamma$-rays. Higher values, common to many
predictions of hadron acceleration in AGN jets \citep[see for example][]
{biermann87} do not significantly
increase the $>10$ TeV emission, due to the absorption of higher
energy photons on the EBL. 

The spectrum of the injected secondary electrons is calculated in a
similar way to that of $\gamma$-rays. The time evolution of the
secondary electron spectrum is followed in small time steps accounting
for synchrotron losses and injection via $p-p$ interactions.

The estimation of the quantity $\zeta(E_{p})$ introduced above is key to
evaluating the fluxes from these objects. $\zeta(E_{p})$ depends on
the energy dependent transport of CRs and on the density profile of
the bubbles. An estimation of the density profile in turn requires an
estimation of the fraction of the bubble pressure provided by CRs
($X_{CR}$) and hence the remaining pressure provided by hot gas.
The left panel of Fig~\ref{f2} shows the relative expected X-ray flux (from inside the
bubbles) and the total $\gamma$-ray flux (from inside and outside the
bubbles) as a function of $X_{CR}$, for two different assumed internal
temperatures. X-ray fluxes above the dashed horizontal line are
effectively excluded by the observed X-ray contrast of the bubbles.
The CR spatial distribution is derived by a numerical transport
simulation of energy dependent diffusion assuming the value suggested
by \citet{voelk96} eq. 9, combined with advection.  Note
that for the purpose of the diffusion calculation a magnetic field
inside and outside the bubbles was assumed to be the same. In general
the diffusion inside the bubbles is much more uncertain and could have
a significant effect on the observed $\gamma$-ray flux.  The right-hand panel 
of Fig~\ref{f2}
shows (dashed line) the fraction of CRs confined within the bubble
under these assumptions. Integrating over this curve, the total CR
energy inside the bubbles is 74\%. The remaining curves on
Fig~\ref{f2} (right) show $\zeta(E_{p})$ calculated numerically for the two
cases marked on Fig~\ref{f2} (left): $X_{CR}=0.5$,
$T_{\mathrm{int}}=T_{\mathrm{ext}}$ (dash-dotted curve, solid circles
in the left panel of Fig~\ref{f2}) and $X_{CR}=1.0$ (solid curve, indicated by open
squares in the left panel Fig~\ref{f2}). For these calculations we assume and
external density profile based on X-ray measurements 
\citep[see Fig. 4 of][]{nulsen05b} and bubble rims with a compression
ratio of 4. It can be seen that at high energies ($>$ TeV) the results
are rather insensitive to the internal density of the bubbles, as most
emission arises from the regions just outside the bubbles.
 
\begin{figure*}
\centering
\vspace{-1mm}
\includegraphics[width=17.5cm]{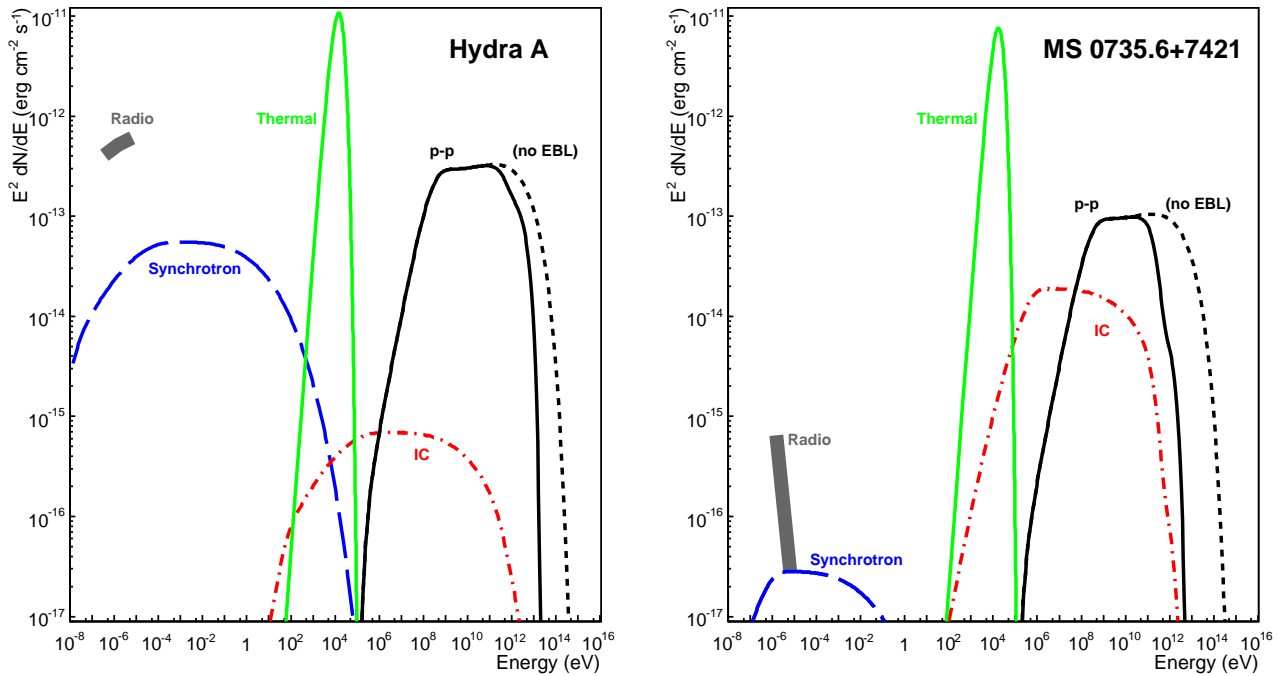}
\vspace{-6mm}
\caption{ Model spectral energy distributions for Hydra A (left) and
  MS\,0735.6+7421 (right).  $\gamma$-ray emission arising from the
  decay of pions produced in hadronic interactions, is shown with
  (solid line) and without (dashed line) the effect of EBL
  absorption. The inverse Compton and Synchrotron radiation of
  secondary electrons and positrons are calculated assuming magnetic
  field strengths of 30$\mu$G and 0.13 $\mu$G for Hydra A and
  MS\,0735.6+7421, respectively. The CMB is assumed to the dominant
  target photon field for inverse Compton scattering. Radio data are
  taken from Lane et al (2004) and Cohen et al (2005).}
\label{f3}
\end{figure*}

\section{Results \& Discussion}

\subsection{Expected broad band emission}

\begin{figure}
\centering
\vspace{-1mm}
\includegraphics[width=8.6cm]{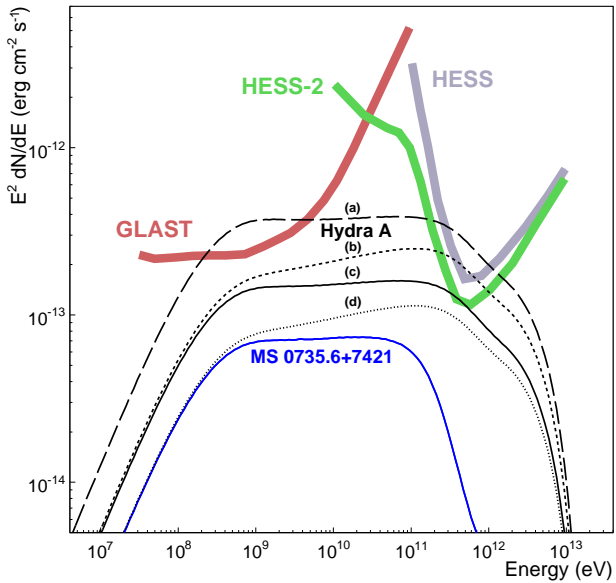}
\vspace{-5mm}
\caption{ Predicted $\gamma$-ray emission for the most powerful known
  AGN outbursts. Model curves are compared to the nominal
  sensitivities of the GLAST, HESS and HESS-2
  $\gamma$-ray detectors, for 5 year and 50 hour observations, respectively. 
  For Hydra A four curves are shown: (a) $X_{CR}=1.0, \eta=1$ implying very cold
  material inside the bubble, (b)$X_{CR}=1.0$ with no gas within the bubbles,
 (c) $X_{CR}=0.5, \eta=2$ and (d) as for (b) but with $PV = 4.1 \times 10^{60}$ erg
 rather than the $9 \times 10^{60}$ ergs used for the other curves.
}
\label{f4}
\end{figure}

Fig.~\ref{f3} shows the resulting broad-band spectral energy
distributions calculated for Hydra A and
MS\,0735.6+7421. Fig.~\ref{f4} compares the expected $\gamma$-ray
emission for the three outbursts considered here.  The predicted HE
and VHE $\gamma$-ray flux of Hydra A is the highest of the three
clusters.  The larger energy injected in CRs in the other two systems
does not compensate for their larger distance (see Fig. \ref{f4}). The
last column of table~\ref{table1} shows the flux expected around
10~GeV ($F^{\prime}_{\nu}$) under the assumption that the total energy
in CR hadrons is equal to $3PV$, i.e. $X_{CR}=1$.  For Hydra A, four
curves are shown, for different assumptions on the distribution of
target material and the energy in cosmic rays:
(a)$X_{CR}=1.0, \eta=1$ implying very cold material inside the bubbles, 
(b)$X_{CR}=1.0$ with no gas within the bubbles,
(c)$X_{CR}=0.5, \eta=2$ and (d) as for (b) but with $PV = 4.1 \times 10^{60}$ erg
 rather than the $9 \times 10^{60}$ ergs used for the other curves.
The two estimates for $PV$ in Hydra-A come from different approaches, 
a shock model \citep{nulsen05b} and summing the $PV$
contributions of individual bubble components \citep{wise07}
and can both be considered valid estimates of the outburst energetics.
We find that EBL absorption in the
case of Hydra A becomes important above 1 TeV. In the most distant
system MS\,0735.6+7421 EBL absorption is already severe above about
200~GeV and the expected $\gamma$-ray flux drops dramatically at
higher energies. 

Secondary electrons resulting from the p-p collisions will lead to the
production of electromagnetic signals via interactions with magnetic
and radiation fields.  For computing the observational signature
associated with secondary electrons in Hydra A we use a magnetic
field of 30 $\mu$G \citep{taylor93} and inverse Compton
up-scattering by the CMBR only. We note that the magnetic field of
relevance here is that in the regions where most target material
exists, i.e. just outside the bubbles.  We can compare the expected
synchrotron emission in the radio band with the observed level of
radio flux \citep{lane04} since in
principle our model could be tested by the expected synchrotron signal
from secondary electrons. However, the observed radio emission seems
to be associated with primary electrons; exceeding the expectation for
secondaries by a factor $\sim50$. In the 2 -- 10 keV X-ray band the
thermal emission of the Hydra A lobes, $\sim 10^{-11}$ erg cm$^{-2}$
s$^{-1}$ \citep{edge92} exceeds our predicted secondary
synchrotron emission by 3 orders of magnitude (see the left-hand panel
of Fig.~\ref{f3}). It therefore appears that in the case of Hydra A
the emission of secondary particles is completely buried by other
processes and that the GeV-TeV $\gamma$-ray range may be the only
wavelength band in which the existence of energetic hadrons can be
probed.

The situation is somewhat different for MS\,0735.6+7421.  The observed
radio spectrum is very steep, particularly in the outer lobes 
\citep{cohen05}. This means that although the low frequency
emission is likely dominated by primary electrons the observed radio
flux of the lobes at $\sim$ 1 GHz is very low ($\sim 2$ mJ) - of the
same order as the predicted radio emission from secondary electrons.
We find that under the assumption that the total energy in hadronic
CRs is equal to $3PV$ that a $B$-field of $> 130$ nG would produce a
high frequency hardening of the radio spectrum which is not observed
(see the right-hand panel of Fig.~\ref{f3}). A value of $\sim$130 nG
(the upper limit in this scenario) is rather low for the central
region of a cluster and lies 3 orders of magnitude below the value of
equipartition with relativistic particles \citep[estimated by][as 100 $\mu$G]{mcnamara05}. 
 We note that radio observations of
this object at $>$1.4 GHz are highly desirable to probe the existence
of secondary electrons (and hence CR hadrons) in this object.

\subsection{$\gamma$-ray observability of these objects}

It is clear from the above discussion that galaxy clusters which
harbour extraordinarily powerful AGNs are at the edge of
detectability with current and near future $\gamma$-ray instruments
if the observed bubbles are dominated by the pressure of 
relativistic particles ($X_{\rm CR} \simeq 1$).
Hydra A is the closest cluster-scale AGN outburst known and appears to be the most promising
target of this kind for $\gamma$-ray observations. Furthermore, its
emission is only moderately affected by EBL absorption in the VHE
$\gamma$-ray regime. Our calculations show that, depending on the
detailed properties of the source, it may be detectable using the
currently operating HESS instrument and the upcoming GLAST
mission. With detections from both instruments it might be possible to
determine the shape of the spectrum of CR protons in this
system.  Additionally, the extended $\gamma$-ray emission expected
from Hydra A could be resolved by instruments such as HESS, but is
not sufficiently extended to significantly degrade the detection
sensitivity with respect to the point-source case.
A $\gamma$-ray detection of Hydra-A would provide a clear 
signature of hadronic CR dominance of the bubble pressure.

The situation is less auspicious for the other two examples of cluster
scale AGN outbursts.  Both systems are more powerful than Hydra A but
their distances make detection of a $\gamma$-ray signal more
difficult. Furthermore, their larger redshifts compared to Hydra A
imply rather severe absorption by the EBL, making a $>$1 TeV detection
of these objects extremely difficult. It is very likely that these
systems can only be detected with the next generation of $\gamma$-ray
telescopes (both ground based and space borne).

\section*{Acknowledgements}

We would like to thank Felix Aharonian, Heinz V\"olk, Olaf Reimer and
Werner Hofmann for very useful discussions, Sven Van Loo for his 
careful reading of the manuscript and an anonymous referee 
for very constructive comments. JAH is supported by a PPARC Advanced Fellowship.

\label{lastpage}

\end{document}